\documentclass{mn2e}
\usepackage[dvips]{graphicx}
\usepackage{amssymb}

\title[Forming IMBHs in Globular Clusters]{Intermediate-Mass Black Holes in Globular Clusters}
\author[Yu-Qing Lou and Yi-Hong Wu]{Yu-Qing Lou$^{1,2,3}$ and Yi-Hong
Wu$^1$\\
    $^1$ Department of Physics and Tsinghua Center for Astrophysics (THCA), Tsinghua University, Beijing
    100084, China; \\
    $^2$ Department of Astronomy and Astrophysics, the University of Chicago, 5640 South Ellis Avenue, Chicago, IL 60637, USA;\\
    $^3$ National Astronomical Observatories, Chinese Academy of Sciences, A20, Datun Road, Beijing, 100021, China}

\date{Accepted 2012 January 26.
    Received 2012 January 7; in original form 2011 September 8}
\pagerange{\pageref{firstpage}--\pageref{lastpage}}\pubyear{}

\def\LaTeX{L\kern-.36em\raise.3ex\hbox{a}\kern-.15em
    T\kern-.1667em\lower.7ex\hbox{E}\kern-.125emX}

\begin{document}

\maketitle

\label{firstpage}

\begin{abstract}
There have been reports of possible detections of
  intermediate-mass black holes (IMBHs) in globular clusters (GCs).
Empirically, there exists a tight correlation between the central
  supermassive black hole (SMBH) mass and the mean velocity
  dispersion of elliptical galaxies, ``pseudobulges" and classical
  bulges of spiral galaxies.
We explore such a possible correlation for IMBHs
 in spherical GCs.
 In our model
  of self-similar general polytropic quasi-static dynamic
  evolution of GCs,
  a criterion of forming an IMBH is proposed.
The key result is $M_{\rm BH}=\mathcal{L}\sigma^{1/(1-n)}$
  where $M_{\rm BH}$ is the IMBH mass, $\sigma$ is the GC
  mean stellar velocity, $\mathcal{L}$ is a
  coefficient, and $2/3<n<1$.
\end{abstract}


\section{Introduction}

For star-forming molecular clouds, collapsed massive dense
  gas cores eventually lead to luminous new-born stars
  burning nuclear fuels.
Analogously but on larger scales, we speculate that grossly
  spherical core-collapses of globular clusters (GCs) could
  also cause something singular around the dense centre;
such central singularities may form
  IMBHs.
Observations of GC cores indicate that the central
  concentrations of nonluminous materials are likely
  due to IMBHs (Bahcall \& Ostriker 1975).
Mass accretions onto IMBHs were proposed to power
  GC ULX sources (e.g. Farrell et al. 2009).

On much larger scales further, observations of galaxies reveal a
  strong correlation between the mass $M_{\rm BH}$ of the central
  SMBHs and the mean velocity dispersion
  $\sigma$ of the host stellar bulge in the form of $\log(M_{\rm
  BH}/M_\odot)=\epsilon+\delta\log(\sigma/\sigma_0)$ where $\epsilon$
  and $\delta$ are two coefficients and $\sigma_0$ is a velocity
  dispersion normalization (e.g. Tremaine et al. 2002).
We naturally expect
  a similar $M_{\rm BH}-\sigma$ power-law relation for GCs.

Evidence for central IMBHs in GCs have been debated
  extensively and their possible existence bears important
  consequences for both the formation and evolution of GCs (e.g.
  Grindlay \& Gursky 1976;
  Maccarone et al. 2007, 2008;
  Zaharijas 2008).
The central brightness excesses observed in several GC cores
 (e.g. Diorgovski \& King 1986) are compatible with the
 presence of central IMBHs.
Properties of IMBHs correlate with various properties of GCs,
  including stellar density profiles, central stellar dynamics,
  luminosities of GCs, the mass-to-light ratios, surface
  brightness, rotation amplitudes, position angles, dark matter
  densities and the mean stellar velocity dispersion $\sigma$
  (Gebhardt et al. 2000; Zheng 2001; Ulvestad et al.
  2007; Zepf et al. 2007; Bash et al. 2008; Noyola et al.
  2008; Zaharijas 2008).
These empirical correlations may shed light on the origin and
  evolution histories of IMBHs and their host GCs.
Among such observed relations, $M_{\rm BH}$
  and $\sigma$ correlate tightly
  (e.g.
  Gebhardt et al.
  2002; Safonova \& Shastri 2010).

Self-similar solutions for general polytropic
  hydrodynamics of a self-gravitating fluid
  with spherical symmetry were constructed recently.
Asymptotic behaviours of novel quasi-static solutions in
  a single polytropic fluid have been revealed by Lou \&
  Wang (2006, 2007) and was used to model rebound (MHD)
  shocks in SNe.
Such solutions were applied to clusters of galaxies
  (Lou et al. 2008) for possible galaxy cluster winds.
In this Letter, we invoke such quasi-static solutions
  to model dynamic evolution of host GCs and formation of
  IMBHs and to establish $M_{\rm BH}-\sigma$ power laws.

Various aspects of GCs have been studied extensively
 (e.g. Benacquista \& Downing 2011 and extensive
 references to excellent reviews therein).
We focus on the quasi-static self-similar GC dynamic evolution
 (say, induced by the gravothermal instability) in the late
 phase of the pre-collapse regime; such asymptotic solution
 for GC evolution leads to diverging mass density at the center.
Physically, as the inner enclosed core mass becomes
 sufficiently high within a radius comparable to its
 Schwarzschild radius, an IMBH forms inevitably (e.g.
 through mergers of stellar mass black holes or runaway
 collisions and coalescence or merging of stars to form
 supermassive stars and to trigger subsequent
 e$^{\pm}$ pair instabilities therein).
After forming such a central IMBH in the core, pertinent
 post-collapse mechanisms continue to operate: e.g., mass
 segregation maintains more massive stars around the central
 IMBH, while the `binary heating' (e.g. Hurley et al. 2007)
 from primordial stellar binaries survived the IMBH
 formation tends to resist further core collapse or
 to drive post-collapse oscillations.
The N-body GC simulations by Hurley et al. (2007) of up
 to $10^5$ stars and initial 5\% binaries may eventually
 reach a GC core binary frequency as high as 40\% at the
 end of the core-collapse phase.
Shown in their figure 3, this core binary frequency
 actually fluctuates between $\sim 10$\% to $\sim 40$\%.
We would expect that an IMBH forms at the GC centre
 rapidly during the core-collapse phase but do not
 know exactly when.
Physically, this IMBH would engulf the central stars
 and binaries at the epoch of IMBH formation.
Such an explosive event may give rise to a powerful
 gamma-ray burst and a shock wave surrounding the
 GC centre.
The slower relaxation, evolution and accretion
 then persist on a much longer time scale.

GCs are close to spherical;
in their dynamic evolution, lumpiness observed could either result
 from merging disturbances and tidal disruptions or provide source
 of fluctuations that may be classified into acoustic modes,
 gravity modes
 and vortical modes
 on larger scales (Lou \& Lian 2011).
These perturbations may be unstable to trigger
 gravothermal instability.
Analogous to nuclear burnings in a star to resist stellar core
 collapse, the `binary heating' from primordial stellar binaries
 may delay core collapse in GCs (e.g. Meylan \& Heggie 1997).
As the source of `binary heating' is exhausted during a
 GC evolution, the inner core collapse is inevitable.
This ultimately leads to the formation of IMBH which
 may accrete materials from immediate environs.

\section{
  GC Model for a $M_{\rm BH}-\sigma$ Power Law}

 We adopt the same hydrodynamic perspective of Lou \& Jiang
 (2008; LJ hereafter) for the large-scale spherical GC
 dynamic evolution.
GCs are smaller and less massive than typical galactic bulges.
The random stellar velocity dispersion and the `binary
 heating' from primordial stellar binaries in the core
 provide an effective pressure against the GC self-gravity.
This may justify our fluid formalism for GC cores.
In contrast to LJ, we emphatically focus on the reported tentative
 candidates of IMBHs and the properties of the host GCs.
Our main goal is to extend the theory of LJ to GCs and examine
 whether IMBHs and properties of GCs can be sensibly fitted with data.
By data comparisons, our results appear encouraging for such a
 physical connection.
Our predictions for IMBHs and host GCs
 can be tested by further observations.
Meanwhile, we also compare with SMBHs
 in their host galactic pseudobulges.
This may hint at a general validity of such a self-similar
 dynamic evolution in spherical self-gravitating systems.
%

%

In spherical polar coordinates $(r,\ \theta,\ \phi)$,
  the nonlinear general polytropic hydrodynamic partial
  differential equations (PDEs) of spherical symmetry
  are PDEs (1)$-$(4) of LJ with the same notations.
The Poisson equation is automatically satisfied.
%
%
%
As the bulk flow of stellar fluid is slow in GCs, we invoke the
  quasi-static self-similar solutions of Lou \& Wang (2006).
We introduce the transformation in the dimensionless independent
  variable $x$,
 \begin{eqnarray}
 r\equiv K^{1/2}t^nx\ ,\quad\ u\equiv K^{1/2}t^{n-1}v(x)\ ,\quad\
 \rho\equiv\frac{\alpha(x)}{4\pi Gt^2}\ ,\nonumber\\
 P\equiv\frac{Kt^{2n-4}\beta(x)}{4\pi G}\ ,
 \qquad M\equiv\frac{K^{3/2}t^{3n-2}m(x)}{(3n-2)G}\ ,
 \label{trans2}
\end{eqnarray}
with $K$ and $n$ being two scaling parameters; here, $u$, $\rho$,
  $P$, and $M$ are radial velocity, mass density, pressure, and
  enclosed mass respectively, while $v(x)$,
  $\alpha(x)$, $\beta(x)$, and $m(x)$ are respectively
  dimensionless reduced speed, mass density, pressure,
  and enclosed mass of $x$ only.

Substituting self-similar transformation (\ref{trans2})
  into nonlinear PDEs (1)$-$(4) of LJ
  and defining $q\equiv 2(n+\gamma-2)/(3n-2)$,
%
 we derive two coupled nonlinear
 ordinary differential equations (ODEs) for $\alpha'$ and $v'$
\begin{eqnarray}
D(x,\alpha,v)\alpha'=N_1(x,\alpha,v),\ \
D(x,\alpha,v)v'=N_2(x,\alpha,v)\ ,\label{ODEs}
\end{eqnarray}
where three functionals $D$, $N_1$ and $N_2$ are defined
 explicitly in Hu \& Lou (2009) with zero magnetic field.
An exact global static solution of eq (\ref{ODEs}) in physical
 dimensions, known as the singular general polytropic sphere, has $u=0$,
\begin{eqnarray}
\rho=\frac{A}{4\pi G}K^{1/n}r^{-2/n}\ ,
 \qquad
M=\frac{nAK^{1/n}}{(3n-2)G}r^{(3n-2)/n}\ ,
\label{staticbeforetrans}
\end{eqnarray}
where coefficient $A\equiv
\big[\frac{n^{2-q}}{2(2-n)(3n-2)}\big]^{-1/(n-3nq/2)}$.
This solution serves as an
 asymptotic `quasi-static' solution for small $x$;
i.e. they are leading terms of $v(x)$ and $\alpha(x)$
 and there exist higher order terms for a self-similar
 asymptotic evolution.

For such quasi-static self-similar hydrodynamic asymptotic
 solutions at small $x$, we consistently presume
$$\alpha=Ax^{-2/n}+Jx^{S-1-2/n}\ \qquad\hbox{ and }
 \qquad v=Lx^S\ ,
 \qquad\qquad\qquad\qquad \label{quasiav}$$
in two coupled nonlinear ODEs (\ref{ODEs}),
and derive two nonlinear algebraic equations
 for the coefficients $J$, $S$ and $L$,
\begin{eqnarray}
n(S-1)J=(S+2-2/n)AL\ ,\label{JSL1}
\end{eqnarray}
\begin{eqnarray}
\bigg[\frac{n^2}{2(3n-2)}+\frac{(3n-2)}{2}{\cal W}\bigg]
\big[S^2+\frac{(3n-4)}{n}S\big]
\nonumber\qquad\\
+\frac{n^2+(3n-2)^2(1-4/n){\cal W}}{(3n-2)}=0\ ,\qquad\label{S}
\end{eqnarray}
where ${\cal W}\equiv n^2q/[2(2-n)(3n-2)]$.
Once the proper roots
  of $S$ are known, coefficients $J$ and $L$ are related
  by eq (\ref{JSL1}); only one is free to choose.
The existence of SPS solution (\ref{staticbeforetrans})
  and the requirement of $\Re (S)>1$ constrain the
  parameter regime of such quasi-static solution.
Oscillatory behaviours can also emerge (Lou \& Wang 2006), which
  may be relevant to the interesting controversy of gravothermal
  and post-collapse oscillations in GCs.
For a sufficiently small ${\cal W}\neq 0$,
  we can have two roots $S>1$ from quadratic eq (\ref{S}).
It also happens for one root $S>1$ and the other root $S<1$.

As a physical requirement for sensible similarity solutions of a
  general polytropic flow, both $v(x)$ and $\alpha(x)$ approach
  zero at large $x$.
Thus for either $x\rightarrow 0^+$ or $x\rightarrow +\infty$, the
  reduced velocity $v\rightarrow 0$, which means at time $t$, for
  either $r\rightarrow 0^+$ or $r\rightarrow +\infty$ the flow
  speed $u\rightarrow 0$, or at a radius $r$, when $t$ is either
  short or long enough, the radial flow speed $u\rightarrow 0$.
This model describes a self-similar dynamic GC evolution
  towards a quasi-static configuration a long time later
  and may grossly fit relaxed spherical GCs.

From the general polytropic EoS (LJ), the pressure is
  $$P=(4\pi)^{\gamma-1}G^{q+\gamma-1}(3n-2)^q
  K^{1-3q/2}{\rho}^{\gamma}M^q$$
  and $\sigma_L(r,\ t)=(\gamma P/\rho)^{1/2}$ is
  the local stellar velocity dispersion in a GC.
Asymptotically as $t\rightarrow +\infty$,
  $\sigma_L(r,\ t)$ becomes
\begin{eqnarray}
\sigma_L(r)=\gamma^{1/2}K^{1/(2n)}n^{q/2}A^{(q+\gamma-1)/2}r^{(n-1)/n}.
\label{sigmal}\end{eqnarray}
To check against data, we derive the spatial average
  of velocity dispersion $\sigma$ in a GC.
The GC boundary is taken as either radius $r_c$ where
  mass density $\rho_c$ is indistinguishable
  from the surrounding or the tidal radius.
Within $r_c$ of a GC, the spatial average of
  stellar velocity dispersion $\sigma_L(r)$ is
\begin{eqnarray}
\sigma=\frac{3}{4\pi {r^3_c}}\int_0^{r_c}\sigma_L(r)4\pi r^2dr=
\mathcal{Q}K^{1/2}\qquad\qquad\qquad\quad\nonumber
\\ \quad\equiv [3n^{1+q/2}{\gamma}^{1/2}/(4n-1)](4\pi
G\rho_c)^{(1-n)/2}A^{3nq/4}K^{1/2}\ .\nonumber
\end{eqnarray}
%
We invoke a heuristic criterion of forming
 an IMBH in a GC (LJ).
An IMBH mass $M_{\rm BH}$ is given by $M_{\rm
  BH}={r_s}c^2/(2G)$ where $r_s$ is its
  Schwarzschild radius and $c$ is the speed of light.
By solution (\ref{staticbeforetrans}) and when
  $$\frac{nAK^{1/n}}{(3n-2)G}r^{(3n-2)/n}=rc^2/(2G)\ ,$$
  an IMBH forms with the Schwarzschild radius
  $$r=r_s=[(3n-2)c^2/(2nAK^{1/n})]^{n/(2n-2)}\ .$$
Only those asymptotic quasi-static GCs with $n<1$
  can thus form central IMBHs (see fig. 1 of LJ).
Consequently,
\begin{eqnarray}
M_{\rm BH}=[c^2/(2G)][(3n-2)c^2/(2nA)]^{n/(2n-2)}K^{1/(2-2n)},
\label{MK}
\end{eqnarray}
or equivalently, the explicitly $M_{\rm BH}-\sigma$ power law
$$M_{\rm BH}=\frac{c^2}{2G}\bigg[
\frac{2nA}{(3n-2)c^2}\bigg]^{n/(2-2n)}\bigg(
\frac{\sigma}{\mathcal{Q}}\bigg)^{1/(1-n)}
\equiv\mathcal{L}\sigma^{1/(1-n)}\ ,$$ where the exponent
  $1/(1-n)>3$ since $2/3<n<1$ (LJ).

To validate our quasi-static model for the nine GCs with available
  observational data and references summarized in Tables 1
  and 3, we fit a $M_{\rm BH}-\sigma$ power law in Fig. 1.
Applying the least-square criterion to the data of Table 1
  (e.g. Meylan \& Mayor 1991; Safonova \& Shastri 2010),
  we obtain
  $M_{\rm BH}=4.1020\times 10^{7}M_{\odot}
  (\sigma/200\hbox{ km s}^{-1})^{3.6251}$ with
  parameters $\{n,\ \gamma,\ \rho_c\}$ being $(0.7241,\
  1.99,\ 4.40 M_\odot {\rm pc}^{-3})$.
This $\rho_c$ value appears somewhat larger but grossly
  consistent with the data in rough orders of magnitude
 (e.g. Meylan 1987). Specifically,
 $\rho_c\sim 0.87 M_\odot\ {\rm pc}^{-3}$ for
  GC 47Tuc (Meylan 1988),
  $\rho_c\sim 0.091 M_\odot\ {\rm pc}^{-3}$ for
  GC NGC6397 (Meylan \& Mayor 1991) and
  $\rho_c\sim 0.42 M_\odot\ {\rm pc}^{-3}$
  for GC G1 (Meylan et al. 2001).
No published $\rho_c$ values for NGC2808,
 M80, M62, NGC6388 and M15 are available.
The nine GCs 47Tuc, NGC2808, $\omega$ Cen, M80, M62,
 NGC6388, NGC6397, M15, and G1 correspond to
 $K=\{1.2,\ 2.0,\ 7.7,\ 1.5,\ 2.1,\ 3.1,\ 0.22,\
 1.9,\ 5.8\}\times 10^{19}\hbox{ cgs unit}$ and
 $r_c=13,\ 16,\ 32,\ 14,\ 17,\ 20,\ 5,\ 16,\
  28\hbox{ pc},$ respectively in rough agreement with
  the estimated data in orders of magnitude.
Harris (1996) reported $r_c$ of 47Tuc, NGC2808, $\omega$ Cen, M62,
  NGC6388, NGC6397, and M15 to be $56,\ 43,\ 88,\ 18,\ 18,\ 11,\
  \hbox{and } 64\hbox{ pc}$, while Bahcall \& Hausman (1976)
  reported $r_c$ of M80 to be $\sim 18\hbox{ pc}$ and Ma et al.
  (2007) estimated $r_c$ of G1 to be $\sim 81\hbox{ pc}$.
\begin{figure}
\includegraphics[width=8cm,height=6cm]{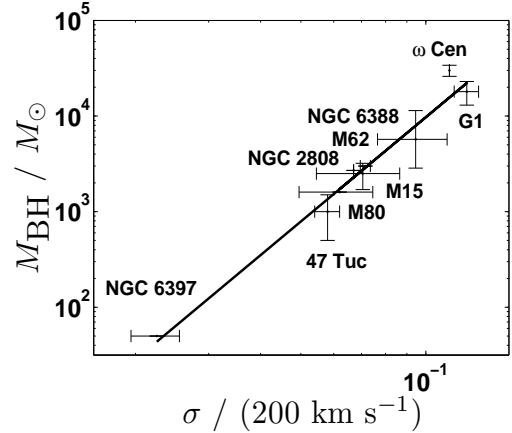}\label{Fig2}
 \caption[]{Mass of IMBH $M_{\rm BH}/M_{\odot}$
 versus mean velocity dispersion $\sigma/(200\hbox{
 km s}^{-1})$ for the nine GCs (Tables 1 and 3).
The solid line is the least-square fit: $M_{\rm BH}
  =4.1020\times 10^{7}M_{\odot}(\sigma/200\hbox{ km
 s}^{-1})^{3.6251}$ with parameters $\{n,\ \gamma,\
 \rho_c\}$ being $(0.7241,\ 1.99,\ 4.40 M_\odot\ {\rm pc}^{-3})$.
}
\end{figure}

%

We now consider a sample of SMBHs in galactic pseudobulges
  (e.g. Kormendy \& Kennicutt 2004) summarized in Table 2.
For the reasons in Hu (2008), we do not regard
  galaxy NGC3227 as containing a pseudobulge.
%
In Table 2, $\sigma_e$ is defined by Gebhardt et al. (2000)
  and Tremaine et al. (2002) as the luminosity-weighted
  rms velocity dispersion within a slit aperture of length
  $2R_e$, where $R_e$ is the effective or half-light radius
  of a galactic bulge.
Parameter $\sigma_8$ is the rms velocity dispersion
  within a circular aperture of radius $R_e/8$.
Ferrarese \& Merritt (2000) used ``central stellar
  velocity dispersion" $\sigma_c$.
There is a relation\footnote{The prime distinguishes this
approximation for $\sigma_8$ from
  the actual value of $\sigma_8$ and the ratio $\sigma'_8/\sigma_8$
  may depend systematically on the velocity dispersion of a galaxy.}
  $\sigma'_8=\sigma_c(8R_{\rm ap}/R_e)^{0.04}$
  (J{\o}rgensen et al. 1995),
  where $R_{\rm ap}\simeq 2''$ (e.g. Davies et al. 1987).
Actually, the difference between $\sigma_e$ and $\sigma_c$
 are much smaller than their errors (e.g. Hu 2008).

\begin{table}\label{tab1}
\caption{Nine GCs with their reported central IMBHs
  and mean stellar velocity dispersions $\sigma$}
\begin{tabular}{@{}lllll}
\hline\hline GC Name&Other Name&$M_{\rm BH}(10^3 M_\odot)$
&$\sigma$(km s$^{-1}$)\\
\hline
NGC104  &47Tuc       &$1.0^{+0.5}_{-0.5}  $&$11.6\pm 0.8$\\
NGC2808 &$--$        &$2.7                $&$13.4       $\\
NGC5139 &$\omega$ Cen&$30\pm 4            $&$22.8       $\\
NGC6093 &M80         &$1.6                $&$12.4\pm 2.5$\\
NGC6266 &M62         &$3.0                $&$14.3\pm 0.4$\\
NGC6388 &$--$        &$5.7^{+5.7}_{-2.85} $&$18.9\pm 3.6$\\
NGC6397 &$--$        &$0.05               $&$4.5 \pm 0.6$\\
NGC7078 &M15         &$2.5^{+0.7}_{-0.8}  $&$14.1\pm 3.2$\\
G1 (M31)&Mayal II    &$18.0^{+5.0}_{-5.0} $&$25.1\pm 1.7$\\
\hline
\end{tabular}
\end{table}

\begin{table}\label{tab2}
\caption{A sample of SMBHs in pseudobulges of galaxies}
\begin{tabular}{@{}llll}
\hline \hline galaxy
         & $M_{\rm BH}(10^8 M_\odot)$
         & $\sigma_e$(km s$^{-1}$)
         & $\sigma_c$(km s$^{-1}$)\\
\hline
NGC1068$^a$  &$0.15                   $&$165\pm 17$&$165\pm 17$\\
NGC2787$^b$  &$0.41^{+0.04}_{-0.05}   $&$210      $&$210      $\\
NGC3079      &$0.025^{+0.025}_{-0.013}$&$146\pm 15$&$146\pm 15$\\
NGC3384$^c$  &$0.16^{+0.01}_{-0.02}   $&$160      $&$160      $\\
NGC3393      &$0.31 \pm 0.02          $&$184\pm 18$&$184\pm 18$\\
Circinus     &$0.011\pm 0.002         $&$ 75\pm 20$&$ 75\pm 20$\\
IC 2560      &$0.029\pm 0.006         $&$137\pm 14$&$137\pm 14$\\
Milky Way$^d$&$0.036\pm 0.003         $&$132.5    $&$132.5    $\\
\hline
\end{tabular}

\medskip
As in Hu (2008), NGC3227 may
  not possess a pseudobulge.

  $^a$ The SMBH mass has been updated by Das et al. (2007).

  $^b$ The stellar velocity dispersion is taken
  from Sarzi et al. (2001).

  $^c$ The stellar velocity dispersion is from
  Busarello et al. (1996).

  $^d$ The SMBH mass has been updated by Falcke et
  al. (2009) and the stellar velocity dispersions
  are from Walter et al. (2006).
\end{table}


A pseudobulge border is at radius $r_c$
  where $\rho$ reaches a value $\rho_c$
  indistinguishable from the environs.
With the least-square fit to Table 2 data,
  we obtain $M_{\rm BH}=2.4350\times 10^{7}M_{\odot}
  (\sigma/200\hbox{ km s}^{-1})^{3.7680}$ with
  parameters $\{n,\ \gamma,\ \rho_c\}$ being $(0.7346,\
  1.999,\ 1.8 M_\odot\ {\rm pc}^{-3})$, respectively.
This estimated $\rho_c$ appears grossly consistent
  with the data in the order of magnitudes, e.g.
  $\rho_c\sim 0.1M_\odot\ {\rm pc}^{-3}$ for the pseudobulge
  in the Milky Way (Lopez-Corredoira et al. 2005).
No published $\rho_c$ values for
  other pseudobulge are available.
The eight pseudobulges NGC1068, NGC2787, NGC3079, NGC3384,
 NGC3393, Circinus, IC2560, Milky Way correspond to values
 of $K=\{3.4,\ 5.8,\ 1.3,\ 3.5,\ 5.0,\ 0.85,\ 1.4,\ 2.6\}\times
 10^{21}\hbox{ cgs unit}$ and $r_c=1.3,\ 1.7,\ 0.79,\ 1.3,\ 1.5,\
  0.64,\ 0.82,\ 1.1\hbox{ kpc},$ respectively in rough agreement
  with the data in orders of magnitudes,
  e.g. Veilleux et al. (1999) reported $r_c$ of NGC3079 to be
  $\sim 3.6\hbox{ kpc}$, while Busarello et al. (1996) estimated
  $r_c$ of NGC3384 to be $\sim 1\hbox{ kpc}$ and Cavichia et al.
  (2011) estimated $r_c$ of Milky Way to be $\sim 1.4\hbox{ kpc}$.

\begin{figure}
\includegraphics[width=8cm,height=6cm]{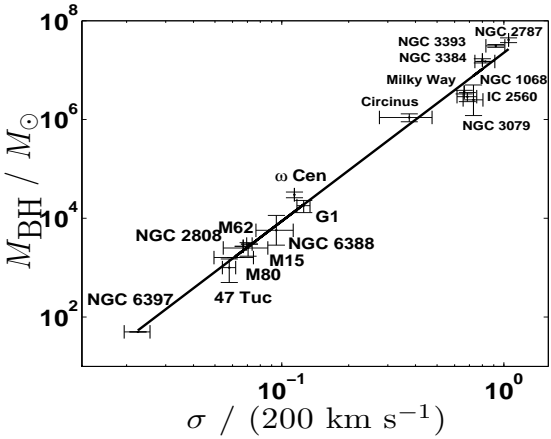}
\label{Fig3}\caption[]{
Mass $M_{\rm BH}/M_{\odot}$ versus
  the mean velocity dispersion $\sigma/\sigma_0$
  ($\sigma_0=200\hbox{ km s}^{-1}$) for galactic
  pseudobulges and nine GCs together in a log-log plot.
The line is the least-square fit to the
  combined data of galactic pseudobulges and nine GCs
  by $\log(M_{\rm BH}/M_{\odot})=7.3529
  +3.4125\log(\sigma/\sigma_0)$ with $n=0.7070$.
}
\end{figure}

Fig. \ref{Fig3} shows the correlation between $M_{\rm BH}$
 and the mean velocity dispersion $\sigma$ of the
 host for samples of galactic pseudobulges
 and of nine GCs.
The joint least-square fit is
$$
\log(M_{\rm BH}/M_{\odot})=7.3529
 +3.4125\log(\sigma/200\hbox{ km s}^{-1})
$$
with $n=0.7070$.
The $n$ value of $M_{\rm BH}-\sigma$ power
 law for nine GCs alone is $0.7241$.
These two $n$ values are close. It seems that
 the $M_{\rm BH}-\sigma$ power law may extend
 down to GCs.

\section{Relation for $M_{\rm BH}-M_{\rm GC}$ power law}

\begin{figure}
\includegraphics[width=8cm,height=6cm]{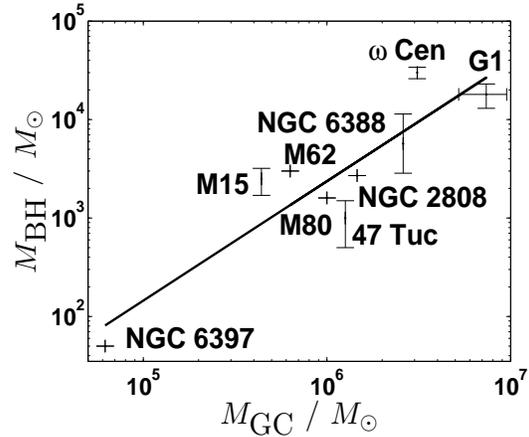}
\label{Fig4}\caption[]{
Masses of central IMBH $M_{\rm BH}$ (in $M_{\odot}$) versus GC
 masses (in $M_{\odot}$) of the nine GCs respectively.
The straight line is the least-square fit to the data in a log-log
display by $M_{\rm BH}/M_{\odot}=1.25\times 10^{-4} (M_{\rm
GC}/M_{\odot})^{1.2128}$ with $n=0.7252$.
}
\end{figure}

By our model analysis, the total mass $M_{\rm GC}$ of a GC is
$${M_{\rm GC}}=
\frac{n(4\pi\rho_c)^{(2-3n)/2}}{(3n-2)}
\bigg(\frac{A}{G}\bigg)^{3n/2}K^{3/2}
\ .$$
%
By eq (\ref{staticbeforetrans}), a smaller $\rho_c$ corresponds
  to a larger $r_c$ and thus a larger $M_{\rm GC}$ with $2/3<n<1$.
By relation (\ref{MK}), $M_{\rm BH}$ and
  $M_{\rm GC}$ are related by the power law
$$
M_{\rm BH}=\bigg[\bigg(\frac{4\pi\rho_c n}{3n-2}\bigg)^{1/3}
\frac{2G}{c^2}\bigg]^{(3n-2)/(2-2n)}{M_{\rm GC}}^{1/(3-3n)}\ .
$$
%
\begin{table} \label{tab3}
\caption{Nine GCs with the reported total GC masses $M_{\rm GC}$}
\begin{tabular}{@{}llll}
\hline\hline GC Name&$M_{\rm GC}(10^6 {\it M_\odot})$&Major Relevant References\\
\hline
47Tuc       &1.26          &Pryor \& Meylan (1993)     \\
NGC2808     &1.46          &Servillat et al. (2008)    \\
$\omega$ Cen&3.1           &Miocchi (2010)             \\
M80         &1.0           &Pryor \& Meylan (1993)     \\
M62         &0.63          &Pryor \& Meylan (1993)     \\
NGC6388     &2.6           &Lanzoni et al. (2007)      \\
NGC6397     &0.062         &Heggie \& Giersz (2009)    \\
M15         &0.44          &van den Bosch et al. (2006)\\
G1 (M31)    &$7.37\pm 2.15$&Ma et al. (2009)           \\
\hline
\end{tabular}
\end{table}
In Fig. \ref{Fig4}, we show the IMBH mass $M_{\rm BH}$ vs
 the GC mass $M_{\rm GC}$.
By the least-square fit to these data,
  we obtain
$$
M_{\rm BH}/M_{\odot}=1.25\times 10^{-4} (M_{\rm
GC}/M_{\odot})^{1.2128}
$$
with $n=0.7252$.
The $\rho_c$ value is $4.93M_\odot\ {\rm pc}^{-3}$,
 grossly consistent with the data in orders of magnitudes.
Note that the $n$ value of $M_{\rm BH}-\sigma$
 power law for the nine GCs is $0.7241$.
These two $n$ values are fairly close, indicating that our
 model may consistently explain both the $M_{\rm BH}-\sigma$
 and $M_{\rm BH}-M_{\rm GC}$ power-laws for the nine GCs.

In Table 4, we list the central SMBH masses and the stellar
  masses of galactic pseudobulges (Hu 2009).
As noted,
  NGC3227 is not taken as a galaxy having a pseudobulge.
Moreover,
  NGC2787 and NGC3384 are two galaxies with composite
  structures consisting of both pseudobulges and small
  inner classical bulges (e.g. Erwin 2008);
we do not treat them as pseudobulges.
  We include the Milky Way as having a pseudobulge with
  $M_{\rm BH}=
  (3.6\pm 0.3)\times 10^6M_{\odot}$ (e.g. Falcke et al. 2009)
  and stellar bulge mass $M_s=(1.3\pm 0.5)\times 10^{10}M_{\odot}$
  (e.g. Dwek et al.1995).
In Table 4, $M_{s,B-V}$ is the pseudobulge stellar mass estimated
  by $K$-band mass-to-light ratio $M/L$ derived from $B-V$ colour.
We do not use the pseudobulge stellar mass calculated by $K$ band
  mass-to-light ratio $M/L$ derived from $r-i$ colour, because
  IC2560, Circinus and NGC3393 lack such data.
Adopting the least-square criterion
 to the data in Tables 4, we obtain
$$
M_{\rm BH}/M_{\odot}=4.2825\times 10^{-8}
 (M_{\rm bulge}/M_{\odot})^{1.3813}
$$
with $n=0.7587$. The $\rho_c$ value
 is $0.08M_\odot\ {\rm pc}^{-3}$.
This estimated $\rho_c$ value appears grossly consistent
 with the data in orders of magnitudes (see Section 2).

\begin{table} \label{tab4}
\caption{A sample of SMBHs in galactic pseudobulges
}
\begin{tabular}{@{}lll}
\hline\hline Galaxies&$\log M_{\rm BH}(+,-)$&$\log M_{s,B-V}$ \\
\hline
NGC1068  &$7.18$&$10.36$ \\
NGC3079  &$6.40\ (0.30,0.30)$&$10.19$ \\
NGC3393  &$7.49\ (0.03,0.03)$&$10.57$ \\
Circinus &$6.04\ (0.07,0.09)$&$9.57$  \\
IC 2560  &$6.46\ (0.08,0.10)$&$10.26$ \\
Milky Way&$6.56$&$10.11$ \\
\hline
\end{tabular}

\medskip
Here, $\log M_{\rm BH}(+,-)$ is the logarithm of
 the mass and $1 \sigma$ error of the SMBH
 and $\log M_{s,B-V}$ is the logarithm of the host
 pseudobulge stellar mass inferred from the $K$-band
 mass-to-light ratio $M/L$ derived from $B-V$ colours.
The SMBH mass and the bulge mass of NGC 1068
 are from Das et al. (2007).
The SMBH mass of Milky Way is from
 Falcke et al. (2009).
\end{table}

\section{Conclusions and discussion}

In summary, for a self-similar quasi-static spherical dynamic
 evolution of a general polytropic GC (LJ), our model analysis
 shows $M_{BH}=\mathcal{L}\sigma^{1/(1-n)}$ and
 $M_{\rm BH}=[(3n-2)/(4\pi\rho_c n)]^{(3n-2)/(6n-6)}
 [c^2/(2G)]^{(3n-2)/(2n-2)}{M_{\rm GC}}^{1/(3-3n)}$
 with $2/3<n<1$.
They agree grossly with current data.

First, we have the exponent $1/(1-n)>3$. Secondly,
 $n$ is independent of uncertainties in estimates
 of $r_c$ and $\rho_c$.
Thirdly, the tight correlation supports a causal
 connection between the formation and evolution
 of an IMBH and the dynamics of host GC.
Finally, while forming an IMBH at the GC centre,
 the spherical general polytropic GC relaxes in
 a self-similar quasi-static phase for a fairly
 long lapse.

Index $n$ holds the key in our
 self-similar quasi-static GC model.
As $\rho=AK^{1/n}r^{-2/n}/{(4\pi G)}$ and $M_{\rm
BH}=\mathcal{L}\sigma^{1/(1-n)}$,
 the smaller the value of $n$ is, the steeper the density profile
 becomes and the smaller the exponent $1/(1-n)>3$ of the
 $M_{\rm BH}-\sigma$ relation is.
When density profile is less steeper,
 there are more materials around the centre
 and the accretion is more effective.
Then it would be more effective to form IMBHs given
 that IMBHs are formed by the collapse of mostly
 stars towards the centre.
The outcome is that for a certain $\sigma$, the
 smaller the mass of an initially formed IMBH is,
 the steeper the density profile is and the smaller
 the $n$.

We conclude that GCs with IMBHs and pseudobulges with
 SMBHs might share qualitatively similar $M_{\rm BH}-\sigma$
 and $M_{\rm BH}-M_{\rm GC}$ power-law relations in general.
These results would bear significance for our theoretical
 understanding of the dynamic evolution of GCs, the
 formation of IMBHs, the connection between GC and
 pseudobulge formation.

\section*{Acknowledgments}
This work was supported in part by THCA,
 by the National Natural Science Foundation of
 China grants 10373009, 10533020, and 11073014,
by the MOST grant 2012CB821800,
by the Tsinghua Univ. Initiative Scientific Research Program,
 and by the Yangtze Endowment and the SRFDP
 20050003088, 200800030071 and 20110002110008,
 from the MoE
 at Tsinghua U.
%

\bigskip

\label{lastpage}
\end{document}